 \documentclass[final,5p,times,twocolumn,10pt]{elsarticle}

\usepackage{amssymb}
\usepackage{color}

\newcommand{\beqn}{\begin{eqnarray}}
\newcommand{\eeqn}{\end{eqnarray}}
\newcommand{\be}{\begin{equation}}
\newcommand{\ee}{\end{equation}}

\newcommand{\mathsym}[1]{{}}

\def \g{\tilde{g}}

\def \w{\widetilde{N}_1}

\begin{document}

\begin{frontmatter}


\title{Dark Matter as a Guide Toward a Light  Gluino at the LHC  }


\author{ Daniel Feldman$^{a}$, Gordon Kane$^{a}$, Ran Lu$^{a}$, and Brent D. Nelson$^{b}$}

\address{$^{a}$Michigan Center for Theoretical Physics,\\
University of Michigan, Ann Arbor, MI 48109, USA \\
$^{b}$Northeastern University, Dept. of Physics,
Boston, MA, 02115, USA}

\begin{abstract}
Motivated by  specific connections to  dark matter signatures,  
we study the prospects of observing the presence of a relatively 
light gluino whose mass is in the range $\sim (500-900)~\rm GeV$  
with a wino-like lightest supersymmetric particle  with mass in
the range of $\sim (170-210)~\rm GeV$. The light gaugino spectra studied
here is generally different from other models, and in particular those with
a wino dominated LSP, in that here the gluinos can be significantly
lighter. The positron excess reported by {the} PAMELA satellite data is
accounted for by annihilations of the wino LSP and their relic
abundance can generally be brought near the WMAP constraints due to 
the late decay of a modulus field re-populating the density of relic dark matter.
We also mention  the recent {FERMI photon} constraints on annihilating dark matter
in this class of models and implications for direct detection experiments 
including CDMS and XENON. We study these signatures in models
of supersymmetry with  non-minimal soft breaking terms derived from both string
compactifications and related supergravity models which generally lead to
non-universal gaugino masses. At the LHC, large event rates from the
three-body decays of the gluino in certain parts of the parameter space are found to
give rise to early discovery prospects for the gaugino sector. 
Excess events at the 5 sigma level
can arise with luminosity as low as $\mathcal{O}$(100) $\rm pb^{-1}$ at a
center of mass energy of 10 TeV and $\lesssim \mathcal{O}$(1) $\rm fb^{-1}$ at  $\sqrt{s} = 7 ~{\rm TeV}$.
 \end{abstract}





\end{frontmatter}

\section{Introduction}
Production of superpartners at the Large Hadron Collider (LHC), the event rates
observed in the recent cosmic ray data \cite{PAM,FERMI}, and potentially in
dark matter direct detection experiments \cite{DD}, all may be linked
to the composition of dark matter. A well motivated class of candidate
models that can be tested on all of these fronts arises when the dark matter
is the lightest R-parity odd supersymmetric particle (LSP)
and has a substantial wino component.

However, even if the LSP is produced at the LHC at a relatively light mass,
a discoverable signal at the LHC may be difficult unless colored superpartners such as the
  gluino, are light. There are  few top down models which generically imply
an LSP that is dominantly wino with a light gluino. The breaking of
supersymmetry (SUSY) through a pure anomaly mediated contribution does
predict an LSP that is a wino, but needs an extension to provide a
consistent model.  String frameworks of interest which give rise to
a light gluino and predict a wino LSP include those based on the
fluxless sector of $G_2$ compactifications from which a realistic
model of soft SUSY breaking has been constructed
\cite{Acharya:2007rc}.  Related models of soft breaking based on the
heterotic string can also yield  a light gluino with a wino-like LSP
\cite{Kane:2002qp,Gaillard:2007jr}, and a light gluino/wino-like
LSP system can also arise from  gaugino mass non-universality in 
 D-brane models of soft SUSY breaking \cite{Corsetti:2000yq}.

That a wino-like LSP is consistent with the PAMELA  satellite data with a mass of order 200 GeV
{has been emphasized} \cite{Pierce,Hisano:2008ti,Kane:2009if,Feldman:2009wv},
and recent works have begun to study the implications of a light wino with a
correspondingly  light gluino \cite{Feldman:2009wv,NUWISC}.
Generally,  for a pure wino, or wino-like dark matter candidate,
the predictions on the relic abundance can be in the vicinity of the WMAP data \cite{WMAP}.
Relatively light wino-like dark matter can 
produce the correct relic density and have a thermal history 
provided it contains non-negligible bino and Higgsino components in extended theories \cite{Feldman:2009wv}, while
the pure wino can do so in a non-thermal paradigm due to the decay of  heavy moduli \cite{MoroiRandall,Acharya:2008bk}
(for recent related work see \cite{Endo,GelminiGondolo,Acharya:2009zt}). The heavy moduli can add large additional post freeze-out entropy to
the primordial particle density from the moduli decay into the SUSY sector. This decay also leads to a release
of winos which annihilate at a temperature much lower than freeze-out.

For the class of models we are interested in here, the gluino mass
is rather light, as dictated by the soft breaking of supersymmetry
and electroweak symmetry breaking, leading to large gluino
production cross sections with subsequent decays of the gluinos via
three-body decay chains. Thus a prominent LHC  signal arises from
multijet production. Some early SUSY discovery prospects in
multijets at the LHC over a broad class of models have been given in
\cite{Hubisz:2008gg}\cite{Randall:2008rw}  (for reviews see
\cite{Kane:2008zz},\cite{BSMLHC}).

Distinctively, here we emphasize well motivated models that yield dark matter
annihilation cross sections consistent with the recent PAMELA data, and
also lead to a spectrum with a light gluino.
The associated production of gluinos
 and a wino-like LSP  lead to a simultaneous probe of supersymmetry at colliders and in present dark matter experiments,
where the gluino is linked to the chargino and neutralino  through its dominant three body decay channels.
The
analysis of electroweak gauginos at colliders with mass degeneracy
between the LSP and chargino has been studied in great detail (for
early work see \cite{Chen:1996ap} and for a recent analysis see
\cite{Buckley:2009kv}) where the soft decays of the chargino  can
lead to a wino LSP and a charged pion, giving rise to a displaced
vertex of a track length of a few centimeters.

The organization of this paper is a follows: 
In Section \ref{g2} we briefly review a soft breaking sector of interest
which gives rise  
to a light wino and a light gluino and serves to illustrate the effects
of the expected high jet multiplicity from the production of gluinos at the LHC. 
We then discuss the numerical simulations that allow us to make contact
between the theory and the data 
and enable a connection between the 
predictions for the LHC and to possible signals of dark matter. Following this, in Section \ref{lhcmodels}
we analyze the early discovery prospects of such models at the LHC for benchmark models and also for 
a large collection of models. The above is all carried out in the framework of the $G_2$ models with 
a pure wino LSP and a light gluino. 

In Section \ref{GeneralPredictions} we examine a larger class of models in the context of relic density and direct and indirect detection of dark matter. We include  models which deviate from a pure wino, but still have a substantial wino component.
Here, as before, the soft breaking of supersymmetry 
and radiative electroweak symmetry breaking dictate the mass of relatively light gluino
in the models of interest.
We optimistically conclude in Section \ref{conc}.

\section{Soft Breaking with Tree and Anomalous Contributions \label{g2}}
\subsection{General Framework}
The underlying framework we work in is described by ${\cal N}=1$
supergravity.   We first consider the $G_2-\rm MSSM$
\cite{Acharya:2007rc} which has a generalized sector of soft SUSY
breaking derived from  both \cite{Gaillard} a tree level
supergravity contribution and an anomalous
contribution.\footnote{Models dominated by a tree level contribution
to the soft terms are also considered in
Section \ref{GeneralPredictions}. } The soft parameters can be
parametrized at the unification scale as $m_0 = s \cdot m_{3/2}$,
$M_a= f_a \cdot m_{3/2}$, $A_{3} = a_{3}\cdot  m_{3/2} $ where
$m_{3/2} \sim  {\cal O}(10 -100) ~{\rm TeV}$ is the gravitino mass,
$m_0$ is a universal scalar mass, $M_a$ are the gaugino masses, and
 $A_{3}$ are the tri-linear couplings of the third generation.  Here the
parameters $(s,f_a,a_3)$ are functions of the microscopic theory
which are determined entirely from the effective supergravity model.
The soft parameters are well approximated by (for the complete
analytical expressions see \cite{Acharya:2007rc}) $s \sim 1$, $f_a =
f'_a \alpha_G -\epsilon ~\eta$, where $f'_{1,2,3} =(0.35, 0.58,
0.64)$, and $\eta = 1 -\alpha_G \delta$  parametrizes gauge coupling
corrections in the tree level sector of the gaugino masses. The
parameter $\epsilon \sim (0.02 -0.03) $ arises as a consequence of
the hidden sector potential which is responsible for tuning the
cosmological constant to zero. The terms entering for the
tri-linears of the third generation are well approximated by $a_3 \sim (3/2) m_{3/2}$
up to small corrections in the normalized Yukawas and 
the normalized volume $V_7$ of the $G_2$ manifold, the latter of which
enters in the determination of the gravitino mass. 
The ratio of the Higgs
 vacuum expectation values is generically in the range $\tan \beta \sim 1.5-2.0$  as
 $\mu$ and $B$ are both taken to arise from the quadratic term in the Kahler
 potential, and are similar in magnitude. 
Here the
largeness of the gravitino mass decouples the scalars while the
gaugino masses are suppressed relative to the gravitino mass, where
the suppression enters via the volume of hidden sector three cycles.
The physical values of the soft parameters are sensitive to the
precise value of the unified gauge coupling and threshold
corrections. The largeness of the gravitino mass generically drives the $\mu$
term to be order $m_{3/2}$ for electroweak symmetry breaking, which
in turn induces a relatively large self energy correction
\cite{PBMZ} to the electroweak gaugino masses \cite{Acharya:2007rc}.

The models studied in reference \cite{Acharya:2007rc} did not have a solution of the $\mu$ problem, 
but simply assumed that $\mu$  and the associated soft breaking term $B \mu$ arose 
from the quadratic term in the Kahler potential.
When the issues of embedding the Standard Model
 in the $G_2$ manifold are fully understood it could happen that $\mu$ is small by symmetry 
arguments, e.g. U(1) charges force the bilinear term to vanish.  
Then if $\mu$ was of the same order as $M_2$ there would be some higgsino mixture in the $G_2$ wino,
 so the theories would have a wino-like  LSP instead of a pure wino one.  In this paper
 we use ``$G_2"$ to refer to the pure wino case of reference \cite{Acharya:2007rc}.  If the theory developed to
 imply a wino-like LSP the results for describing the PAMELA data and LHC predictions would
 change very little; however predictions on dark matter direct detection would be considerably modified.

\subsection{Analysis \label{sim}}
In this work we have implemented the complete analytical
expressions for  soft breaking terms of the $G_2$-MSSM into {\sffamily SOFTSUSY}\cite{ben}.
The analysis includes the gaugino mass threshold corrections
\cite{PBMZ} with  2 loop scalar corrections, 2-loop
RGEs for the Higgs and gaugino masses, $\mu$, and Yukawa and gauge couplings
 \cite{Martin:1993zk,ben}.
  Branching fractions have been computed with
{\sffamily SUSYHIT} \cite{susyhit} and production of signal and
backgrounds  are generated with {\sffamily  PYTHIA  } \cite{PYTHIA}
and {\sffamily  PGS} \cite{PGS} with the level 1 (L1) triggers
designed to efficiently reproduce CMS specifications
\cite{Ball:2007zza} (for detailed discussions see e.g. \cite{FLN}).
Signal and background have been simulated at $\sqrt{s} = (7,10,14)~~
\rm TeV$ in order to generalize our predictions for preliminary LHC
runs and future operational center of mass energies. Specifically,
SM backgrounds have been generated with QCD multi-jet production due
to light quark flavors,
 heavy flavor jets ($b \bar b$,  $t \bar t$), Drell-Yan,
single $Z/W$ production in association with quarks and gluons ($Z$+
jets / $W$+ jets), and  $ZZ$, $WZ$, $WW$ pair production resulting
in multi-leptonic backgrounds. Laboriously, samples were generated
at $\sqrt{s} = (7,10,14) \rm TeV$ with up to 5~$\rm fb^{-1}$ of
luminosity. In {\sffamily  PGS4} jets are defined through a
cluster-based  algorithm which has a heavy flavor tagging efficiency
based on the parametrizations of the CDF Run II tight/loose
(secondary) vertex b-tagging algorithm \cite{CDF2}. The standard
criteria for the discovery limit of new signals is that the SUSY
signals should exceed either $5\sqrt{N_{\rm SM}}$ or 10 whichever is
larger,  i.e., ${N^c_{\rm SUSY}}>{\rm Max}\left\{5\sqrt{N^c_{\rm
SM}},10\right\}$, were $c$ indicates the channel of interest.

 \begin{figure*}[t]
\begin{center}
\includegraphics[width = 9cm , height= 7cm]{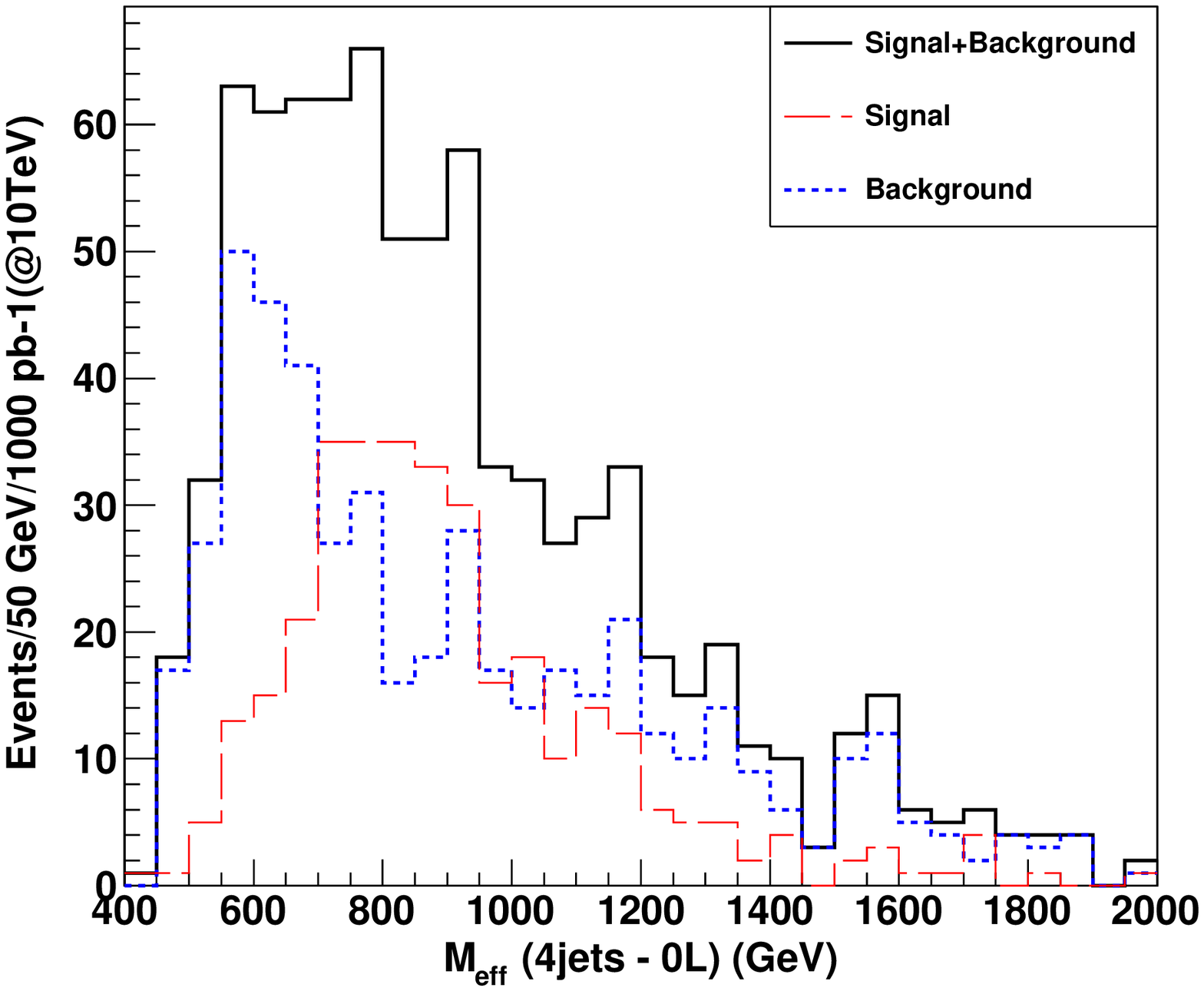}
\includegraphics[width = 9cm , height= 7cm]{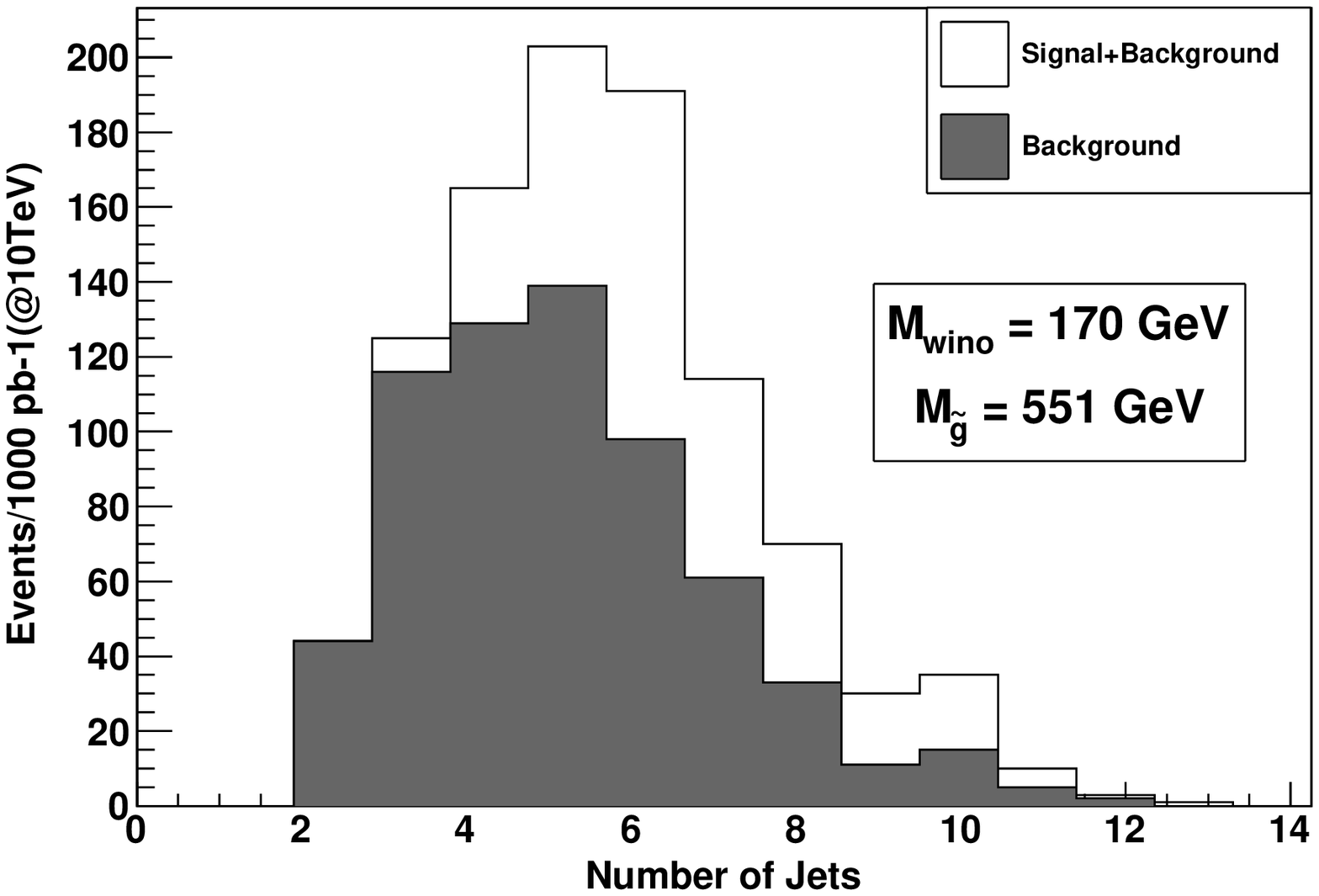}
\includegraphics[width = 9cm , height= 7cm]{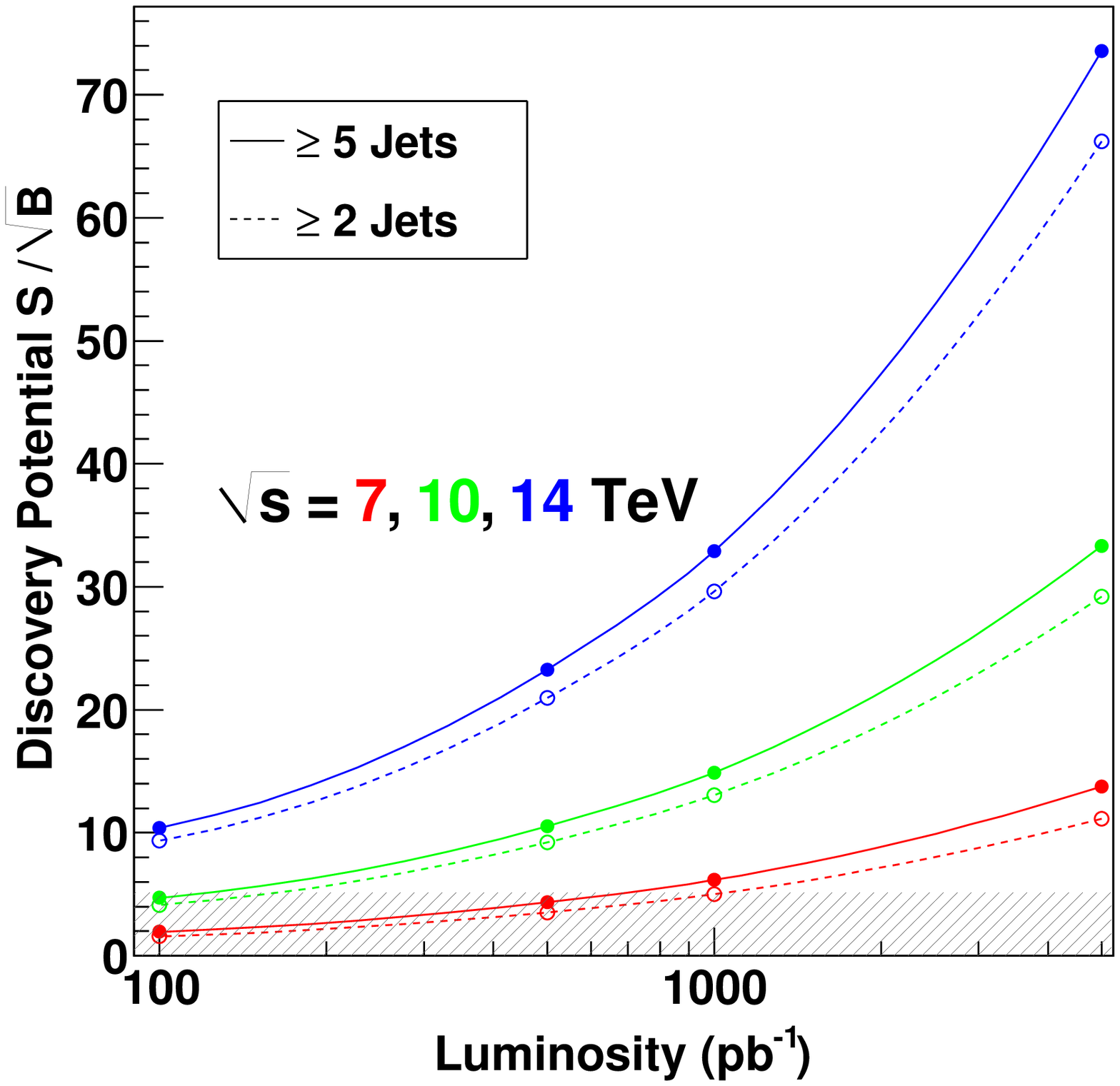}
\includegraphics[width = 9cm , height= 7cm]{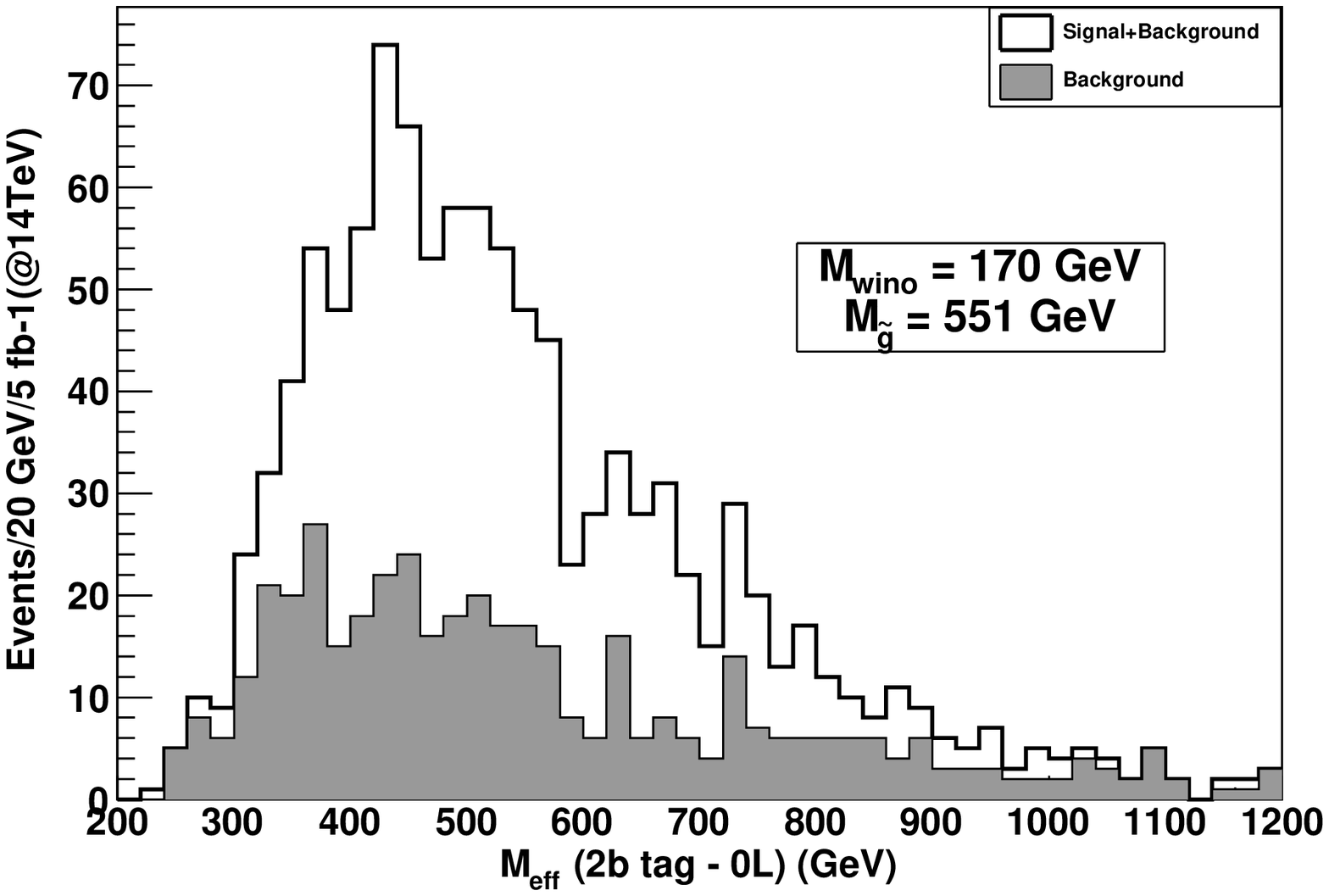}
\caption{ (Color online) Upper left panel:  $M^{4jets}_{eff} =
\sum_{J={1-4}} P^J_T(J) + P^{\rm miss}_T$ at 10 TeV with 1 $\rm fb^{-1}$ for
the $G^1_2$ model benchmark with $S_T \geq 0.25$ (transverse sphericity), $ P^{\rm miss}_T \geq 200 \rm ~GeV$
and a lepton veto. The backgrounds mainly comes from dijets, $t \bar t$ and $W$+ jets. Upper right panel:  Distribution of jet number
showing  excesses in events with large jet multiplicities at low luminosity.
Lower left panel: Discovery reach for the same model with $\sqrt{s} = (7,10,14)~ \rm TeV$. Lower right panel: Same model and cuts as the upper panel
for 14 TeV with 5 $\rm fb^{-1}$ in the variable $M^{2b}_{eff} = \sum_{J={1-2}} P^b_T(J) + P^{\rm miss}_T$.}
\label{fig1}
\end{center}
\end{figure*}

The signature space of the models we probe has distinctive dark matter predictions.
The models we consider are dominated by dark matter annihilations into $W^{+}W^{-}$
and can yield a significant flux of cosmic antimatter in the galactic halo
(for early work see \cite{sa1,sa2,sa3,sa4}).
The annihilation cross section receives an enhancement relative to other SUSY modes
since it is  s-wave  and has a relative strength dictated by the SU(2) gauge coupling
and the wino component of the LSP. 
The models are made consistent with the
relic density constraints as will be discussed.
For the analysis of dark matter annihilation cross sections and their resultant fluxes we employ fragmentation
functions from {\sffamily DarkSUSY }\cite{DS} using {\sffamily  PYTHIA.}
In this work we also model cosmic fluxes with {\sffamily GALPROP v50.1p} \cite{Strong:1998pw}.

\section{Early Discovery Prospects and Concrete Signatures \label{lhcmodels}}
In theories with wino LSPs, the dominant LHC production modes are
not strictly those from strongly produced SUSY. The production modes of the wino ($\w$)  and the lightest
chargino ($\widetilde{C}_1$) are competitive with the gluino ($\g$)
production and frequently are larger. However due to the small
splittings (a fraction of a GeV) between the wino and chargino the decay products here are
soft. {Except} for larger gluino masses, we find that most events
that pass the triggers do indeed come from $\g \g$ production,
though  as much as 30\% of the events come  from
electroweak production. Thus the dominant
production modes are $pp \to [(\g \g), (\w \widetilde{C}_1),
(\widetilde{C}^{\pm}_1,\widetilde{C}^{\mp}_1)]$. The decays modes
lead to rich jet and missing {energy} signatures with a sizeable number
of leptons in the final state.  In particular the dominant decays
are as follows: $\g \to [(\widetilde{N}_2 t \bar t),(\w  b \bar
b),(\w  q \bar q), (\widetilde{C}^{-}_1  {\bar b} t + ~{\rm h.c.}),
(\widetilde{C}^{-}_1  {\bar d} u + ~{\rm h.c.)} ]$ with
 secondary decays  $ \widetilde{N}_2 \to \widetilde{C}_1 W^{*} \to (\widetilde{C}_1 l \nu_l), (\widetilde{C}_1 q \bar q')$ and $\widetilde{C}_1 \to \w W^{*} \to (\w l \nu_l),(\w q \bar q') $ with
 tertiary branchings of the produced standard model particles
$t \to W b$ and $W \to [(q {\bar q'}),(l \nu_l)]$.

The models are rather predictive as they typically require no more than 2-3 branchings
to complete SUSY cascades resulting in lepton and jet signatures. While this is a typical signature of SUSY in 
a generic model, it is actually a prediction of the wino branch of the $G_2$ model as electroweak symmetry breaking corners the 
viable parameter space and thus the viable signature space.
The decays of $\tilde{C}_1 \to \w$ and their jet and lepton by-products  will be very soft
yet there can be radiation of gluon from the initial or final state partons that can generate a relatively hard jet.
Thus one can look for a hard monojet and n-jet events  with large missing energy as an early
indication of the production of supersymmetric events at the LHC.
In Table (\ref{parbench}) we illustrate some typical spectra
found in the $G_2$ models for {$m_{\w}  \sim (170-190)~\rm GeV$} (precisely in the mass range pointed
to by the recent PAMELA data [see Sec.(\ref{positrons})] along with the dominant branching ratio of the gluino {given} in {Table (\ref{branch}).}
 
\begin{table}[t]
\scriptsize
\begin{center}
\caption{Benchmark models predicting a light gluino and a LSP  that is a wino with a degenerate chargino  with a light  second neutralino  (which is mostly bino).
The last four columns carry units of GeV. 
}
\begin{tabular}{c|c |c |c |c| c|c|c|c|c|c|c|c }
$G^{m}_2$  &    $m_{3/2}~(\rm{TeV})$   &   $\delta$   &   $V_7$ &  $\tan\beta$  & $m_{ \tilde g}$ &   $m_{ \tilde W}$ & $m_{ \tilde{C}^{\pm}_1}$ & $m_{ \tilde{N}_2}$  \\
\hline
$G^1_2$   &  38.950  & -2.9  & 30.0 &  1.98     & 551   & 170.2 &   170.4   &   260 \\  
$G^2_2$   &  21.186&  -10.0 &  33.0  & 1.41     & 717   & 173.4 &   174.0   &   190   \\  
$G^3_2$   &  20.700 &  -9.3 &  36.0  & 1.57     & 652   & 176.0 &   176.5   &   185  \\  
$G^4_2$   &   20.618  &-9.1  & 30.0 &  1.71     & 632   & 180.9 &   181.3   &   185 \\   
$G^5_2$   &   35.492  &-5.4  & 32.0 &  1.54     & 761   & 190.5 &   190.6   &   263     \\   
\end{tabular}
\label{parbench}
\end{center}
 \end{table}
  For the $G_2$ models, a central prediction is a relatively light gluino
over the range of wino mass that is capable of describing the PAMELA data {as is illustrated in Table (\ref{parbench}).}
 \begin{table}[t]
  \scriptsize
\begin{center}
\caption{Dominant branching ratios of the gluinos.  }
\begin{tabular}{c|c |c |c |c| c|c|c|c|c|c|c|c }
${\mathcal Br}({\tilde g} \to X)$                 & $G^1_2$  & $G^2_2$ & $G^3_2$ & $G^4_2$ &  $G^5_2$ \\
\hline
${\tilde g}\to b \overline b \w $                       &  14.2     & 6.4    &  10.2    &   11.3  &    19.5\\
${\tilde g}\to q \overline q \w $                       &  21.0      &  7.4    &  12.6   &    14.6  &    10.0 \\
${\tilde g} \to t \overline t {\tilde N}_2$             &    -        &  47.6   &  21.8   &    14.5  &   14.6 \\
$ {\tilde g} \to t \overline b \tilde{C}^{-} + h.c.$    &  18.9   & 16.2    &  20.8    &  20.9    &   24.6  \\
${\tilde g}\to q_u {\overline q}_d \tilde{C}^{-}   + h.c $   &   41.5   &  14.6  &  25.2   &    29.0   &   24.9 \\
 \end{tabular}
\label{branch}
\end{center}
 \end{table}
 \begin{table*}[t]
\begin{center}
\caption{Shown is $\sigma_{\rm SUSY} {\rm (fb)}$,
the theoretical cross section before passing through the detector simulation, $\sigma_{\rm eff} {\rm (fb)}$,
the effective cross section after events have passed the L1 triggers with ${\mathcal L} = 1 \rm fb^{-1}$
at  $\sqrt{s}= 10$ TeV. Observable counts in the number of tagged b-jets and {multijets} are also shown
$N(2b),N(4j)$ along with their signal to square root background ratios. The missing energy cut is $\geq$  200  GeV  and
we have imposed a transverse sphericity cut of $S_T \geq 0.25$.  }
 \scriptsize
\begin{tabular}{c|c |c |c |c| c|c|c|c|c|c|c|c }
    $G^{m}_2$  & $\sigma(\g \g)$   (fb) &   $\sigma(\w \widetilde{C}_1)$  (fb)  &   $\sigma(\widetilde{C}^{\pm}_1 \widetilde{C}^{\mp}_1)$  (fb) &   $\sigma_{\rm SUSY }$ (fb)
   & $\sigma_{\rm eff} $ (fb)  & $N(4j)$&  $\frac{N}{\sqrt{B}}|_{4j}$ & $N(2b)$  & $\frac{N}{\sqrt{B}}|_{2b}$  \\   %
   \hline
$G^1_1$   &  1613 & 996 & 301 & 2910 & 1645 & 416 & 13.3 & 37 & 4.7 \\ 
$G^2_2$ & 236 & 970 & 277 & 1484 & 353      & 79  & 2.5 & 22 & 2.8 \\ 
$G^2_3$  & 481 & 903 & 280 & 1665 & 553     & 133 & 4.2 & 37 & 4.7 \\ 
$G^2_4$  & 648 & 877 & 246 & 1773 & 736     & 217 & 7.0 & 32 & 4.1 \\ 
$G^2_5$ & 182 & 696 & 208 & 1087 & 250      & 64  & 2.0 & 10 & 1.2 \\  
 \end{tabular}
\label{lhcdata}
\end{center}
 \end{table*}
  In {Figure~(\ref{fig1})} (left upper panel) one observes that the models can produce
 detectable multi-jet signals even at $\sqrt{s}= 10$ TeV for $\mathcal L \sim  1 \rm fb^{-1}$
 of integrated luminosity under the standard $5\sigma$ discovery reach criteria in the kinematic
 variable  $M^{4jets}_{eff} =
\sum_{J={1-4}} P^J_T(J) + P^{\rm miss}_T$.
 In {Figure~(\ref{fig1})} (right upper panel) we show the large number of multijet signals.
 The analysis shows that
the model can produce a large excess in hadronic jets over the backgrounds.
The large jet multiplicity arises from the  three body decay of the gluinos
and from jets arising from initial state radiation. We find the discovery limit
is optimal for  4-5 jets with a lepton veto and large missing energy cut.
 The lower right panel exhibits $M^{2b}_{eff} = \sum_{J={1-2}} P^b_T(J) + P^{\rm miss}_T$
with larger luminosity.
The lower left panel shows the discovery reach for the
same model with   $\sqrt{s} = (7,10,14) ~\rm TeV$ and $5\sigma$ can be reached with several hundred
inverse picobarns of data.

\subsection{Global Analysis and Discovery Prospects of Early SUSY \label{early}}
\begin{figure}[t]
\begin{center}
\includegraphics[width = 9cm , height= 6cm]{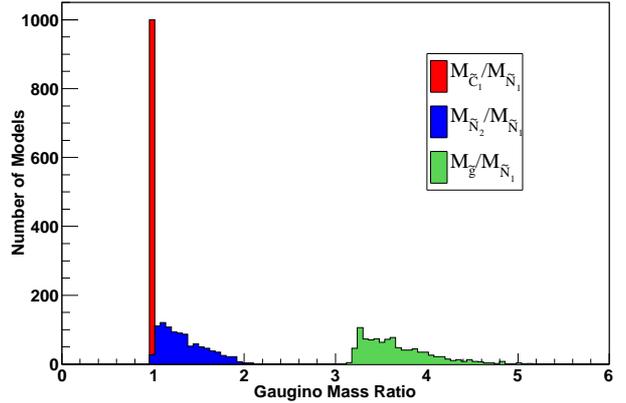}
\caption{ (Color online) Ratio of gaugino masses in the $G_2$ model. The predicted
ratios can be quite different than those that arise in other models
of soft SUSY breaking (for a comparison see Ref. \cite{gnlsp}). The mass
range here for the wino is (170 - 210) GeV
and the gluino lies in the range (500-900) GeV. }
\label{figrat}
\end{center}
\end{figure}

Having established that the highly constrained, and therefore
predictive $G_2$ model can give rise to detectable signals of SUSY
with early LHC data (see also \cite{sekmen}), we now extend the
analysis to a larger region of the $G_2$ parameter space rather than focusing
on a benchmark model. We have performed a
detailed scan of the parameter space of these models over the parameters discussed
in Section \ref{g2},  consistent with
radiative electroweak symmetry breaking subject to the constraint
that the  wino mass is in the range (170 -
210) GeV . We uncover a large parameter space where the gluino can
be relatively light in the $G_2$ model. The majority of the models have a gluino in
the mass range of 500 to 900 GeV (see Fig.(\ref{figrat}) {for the corresponding gaugino mass ratios}).
LHC predictions with light gluino  have been studied recently
\cite{Kuflik,gnlsp,Bhattacharya:2009ij,Baer}, but without
considering the connection to the PAMELA data, which we pursue in the next section.

 \begin{figure}[t]
\begin{center}
\includegraphics[width = 8.75cm , height= 8cm]{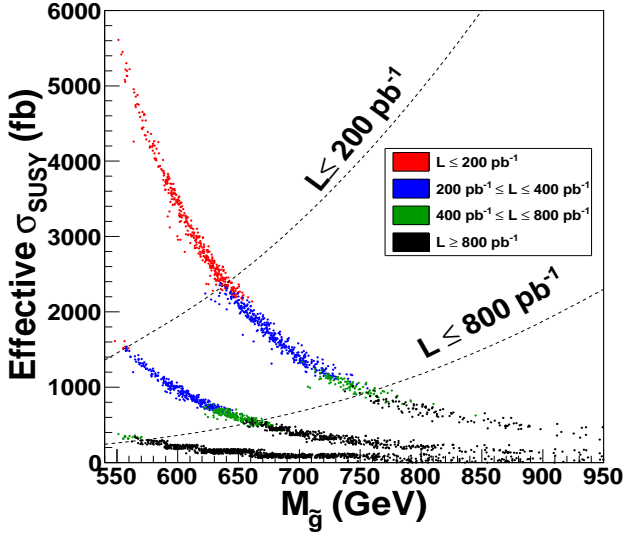}
\caption{ (Color online) Shown is the discovery potential for the gluino at low luminosity and variable LHC center of mass energy, $\sqrt{s} = (7,10,14 )~\rm TeV$ in terms of the effective SUSY cross section (cross section after cuts).
The colored regions are the reach in steps of 200 $\rm pb^{-1}$ (see legend), while the approximated dashed curves are shown for the purpose of illustration.
The missing energy cut is 200 GeV and $S_T \geq 0.25$.
Scanning over optimal signatures, the best channels are 0L+ njets and nbjets. The analysis shows that many of the models
can be discovered at $\sqrt{s}$ = 10 TeV  with order 100 $\rm pb^{-1}$ of luminosity, and that the LHC will be able
to probe a 550 GeV gluino even at $\sqrt{s}$ = 7 TeV with as little as 500 $\rm pb^{-1}$ of luminosity.
}
\label{fig3}
\end{center}
\end{figure}

In {Table (\ref{lhcdata})} we display the relatively large total
theoretical production cross section before cuts ($\sigma_{\rm
SUSY}$ from gluino, neutralino, chargino production) and the
effective SUSY cross section $\sigma_{\rm eff}$ (cross section after
the L1 triggers have been passed).  One observes that the L1
triggers are {well} optimized for these events as a large fraction of the
SUSY cross section is maintained. The substantial missing energy
arises in many of the models from the prompt branching of the gluino
into 2 jets and the LSP wino. Event rates at the LHC are shown in
the 4-jet channel and the 2b channel with just 1 fb$^{-1}$ of
integrated luminosity at $\sqrt{s}= 10$ TeV  along with the ratio of the
signal to the square root of the background. These models can be
discovered very early with the LHC and can begin to be probed at $\sqrt{s} = {\rm 7~TeV}.$

{ Figure ({\ref{fig3}}) }
displays the effective SUSY production cross section after cuts
($\sigma_{\rm eff}$) as a function of gluino mass at various center
of mass energies. The (shaded) colored regions are the necessary
luminosity need for a $5 \sigma$ excess in steps of 200 pb$^{-1}$ of
integrated luminosity where we require at least 5 jets and large
missing energy $\geq 200$ GeV. Thus it is apparent from the analysis
that nearly all the models can produce discoverable signals with low
luminosity. Remarkably, we find that with only
$\mathcal{O}(100\,{\rm pb}^{-1})$ of data at $\sqrt s= $ 10 TeV, the
models will produce large jet-based signals which can be discovered
over the SM backgrounds over a part of the parameter
space, even for gluinos as light as 550 GeV with $\mathcal L \sim
500 ~\rm pb ^{-1}$ at $\sqrt{s}= 7$ TeV.

Models with wino-like LSPs, and thus nearly degenerate charginos and
neutralinos, are well known to be be difficult to study \cite{track}.
 The chargino lifetime can be of order a centimeter, and the second
heavier neutralino can even have order tens of GeV splitting (see
Table (\ref{parbench}) for such theory motivated examples). Once a
set of gluino candidates have been identified, an off-line analysis
focused towards the study of the chargino and neutralino states in
the gluino decay products will be necessary.

 \begin{figure}[t]
\begin{center}
\includegraphics[width = 9cm , height= 8cm]{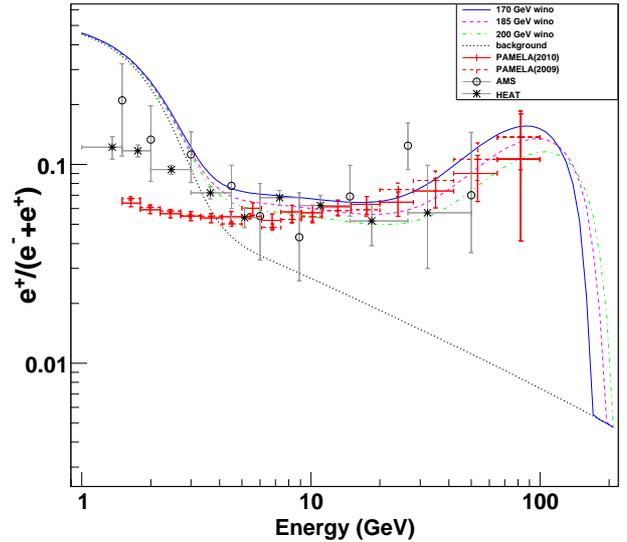}
\caption{(Color online) Rise in the positron fraction predicted from a wino LSP with the PAMELA \cite{PAM}, HEAT and AMS data \cite{heatams}.
Different wino masses are shown to illustrate the range of masses that are well motivated to be a part of a description
of the full Satellite data. Masses somewhat below 170 GeV or a bit above 200 GeV could also provide a reasonable description
of the data.
 }
\label{figpam}
\end{center}
\end{figure}

\section{General Implications of a Wino-Like LSP \label{GeneralPredictions} }
In this section we relax the tight constraints of the $G_2$ theory space
 and explore the possibility of an LSP which has a significant wino
component (``wino-like"), but may also have  non-negligible bino and Higgsino components.
One natural class of models
where such an LSP is achieved are in grand unified models such as $\rm SU(5)$, $\rm SO(10)$, and $\rm E_6$
where the GUT symmetry is broken by a non-singlet $F$ term
leading to gaugino masses at the unification scale that are non-universal, i.e., $M_a = m_{1/2}(1+ \Delta_a)$, $a = 1,2,3$.
Such soft breaking mass terms can give rise to
a wino-like LSP with a light gluino if the high scale values of
the gaugino masses, $M_2$ and $M_3$, are reduced
relative to $M_1$. 
\vspace{-.1cm}
\subsection{Relic Abundance of a Wino-Like LSP}
\vspace{-.1cm}
In a  general setting, the relic density can be equal to the observed one with a wino-like or pure wino LSP
due to the late decay of a modulus field.
 Such is possible in a universe that has a
non-thermal cosmological history \cite{MoroiRandall}.
Thus, for a single heavy modulus field $\Phi$, in the so-called
instantaneous decay approximation
 one obtains a reheat  temperature, $T_R$,
due to the decay $\Gamma_{\Phi}$ by assuming all energy density of $\Phi$ is transferred into radiation.
The modulus decays after freeze-out and the reheat temperature is
$T_R = C^{1/4} \sqrt{{\overline M}_{\rm pl} \Gamma_{\Phi}}$, $C =90/(\pi^2 g_*(T_R))$.
Here $\Gamma_{\Phi} = c_{\Phi} M^3_{\Phi}/{\Lambda^2}$ , $c_{\Phi} \sim 1$, where $ \Lambda \simeq {\overline M}_{\rm pl} \equiv {\overline M}_{\rm pl}/\alpha $,
where $\alpha$ parametrizes deviations from the Planck scale (moduli couplings
at (much lower) intermediate scales have been considered in \cite{Khalil:2002mu,Conlon:2007gk}).
For example, $\alpha = \sqrt{V_7}$ gives $\Lambda = {\overline M}_{\rm pl}/\sqrt{V_7} \sim (2-4) \times 10^{17} {\rm GeV}$
which may be interpreted as an effective string scale.   Under this assumption of non-thermal (NT)
production  one has $\Omega_{\tilde W} \simeq \Omega_{\rm T}|_{T_R}$,
where $\Omega_{\rm T} h^2$ can be computed in the usual manner (see i.e. \cite{Griest:1990kh}).
For the s-wave dominated LSP {interaction}, we obtain $\Omega_{\tilde
W} h^2 \simeq 0.32 \frac{1}{\alpha \sqrt{c_{\Phi}}} (\frac{3*10^{-7}
{\rm GeV}^{-2}}{<\sigma v>})( \frac{m_{\tilde W}}{200 {\rm
GeV}})(\frac{m_{3/2}}{100 {\rm TeV}})^{-3/2}$, where we have used
$m_{\Phi} \lesssim 2 m_{3/2}$, and where $\w \equiv \widetilde W$. A saturation of the error corridor from
the WMAP constraint on $\Omega h^2$
 is then possible for a gravitino in the mass range $(40 -60)$ TeV.
In the $G_2$ models specific calculations of the relic abundance
from moduli decay have been carried out \cite{Acharya:2008bk}
giving a relic density, from a string based construction, a few times larger than the experimental value
unless the gravitino mass is order 100 TeV. 
\begin{figure*}[t]
\begin{center}
\includegraphics[width = 9cm , height= 8cm]{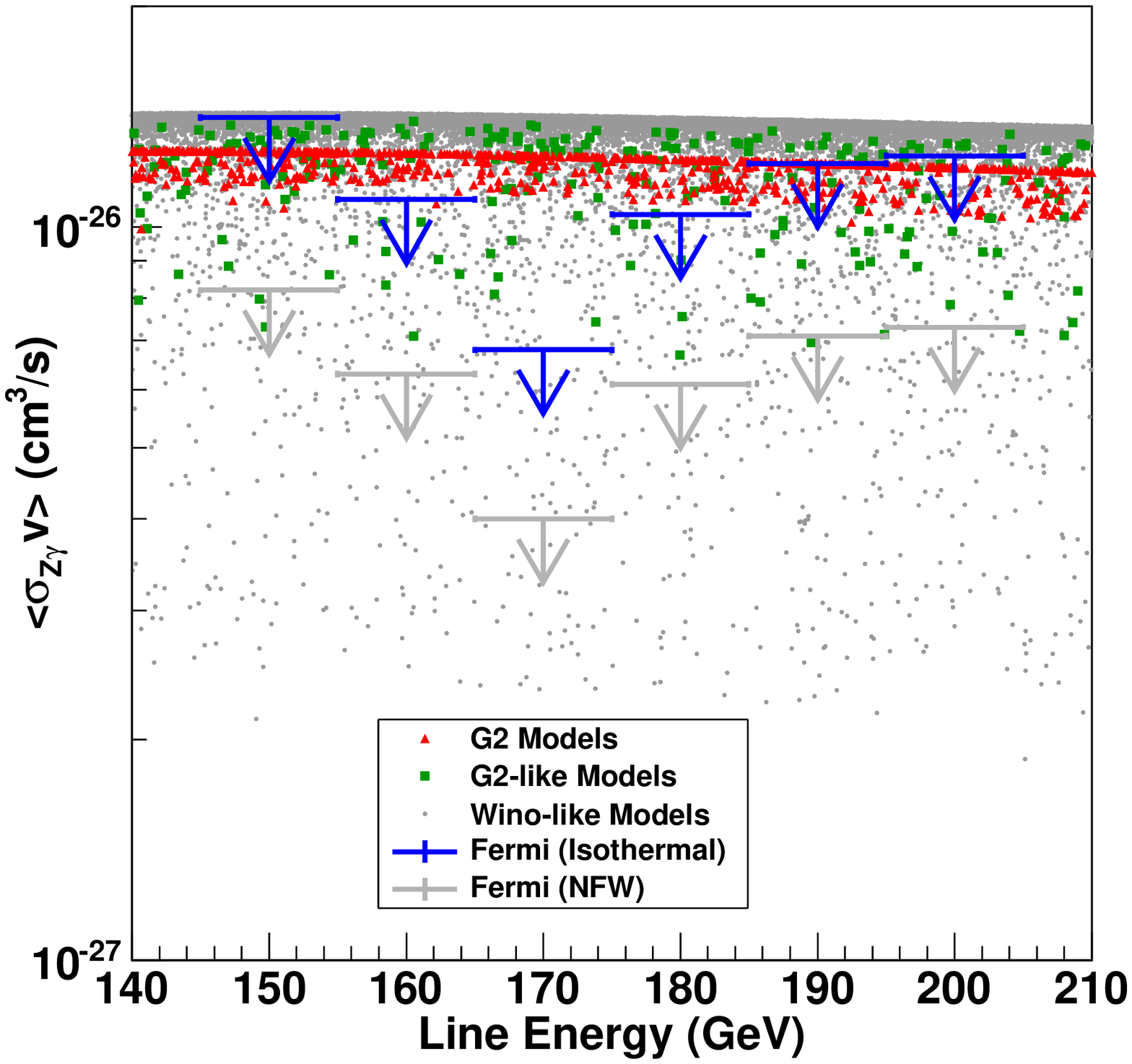}
\includegraphics[width = 9cm , height= 8cm]{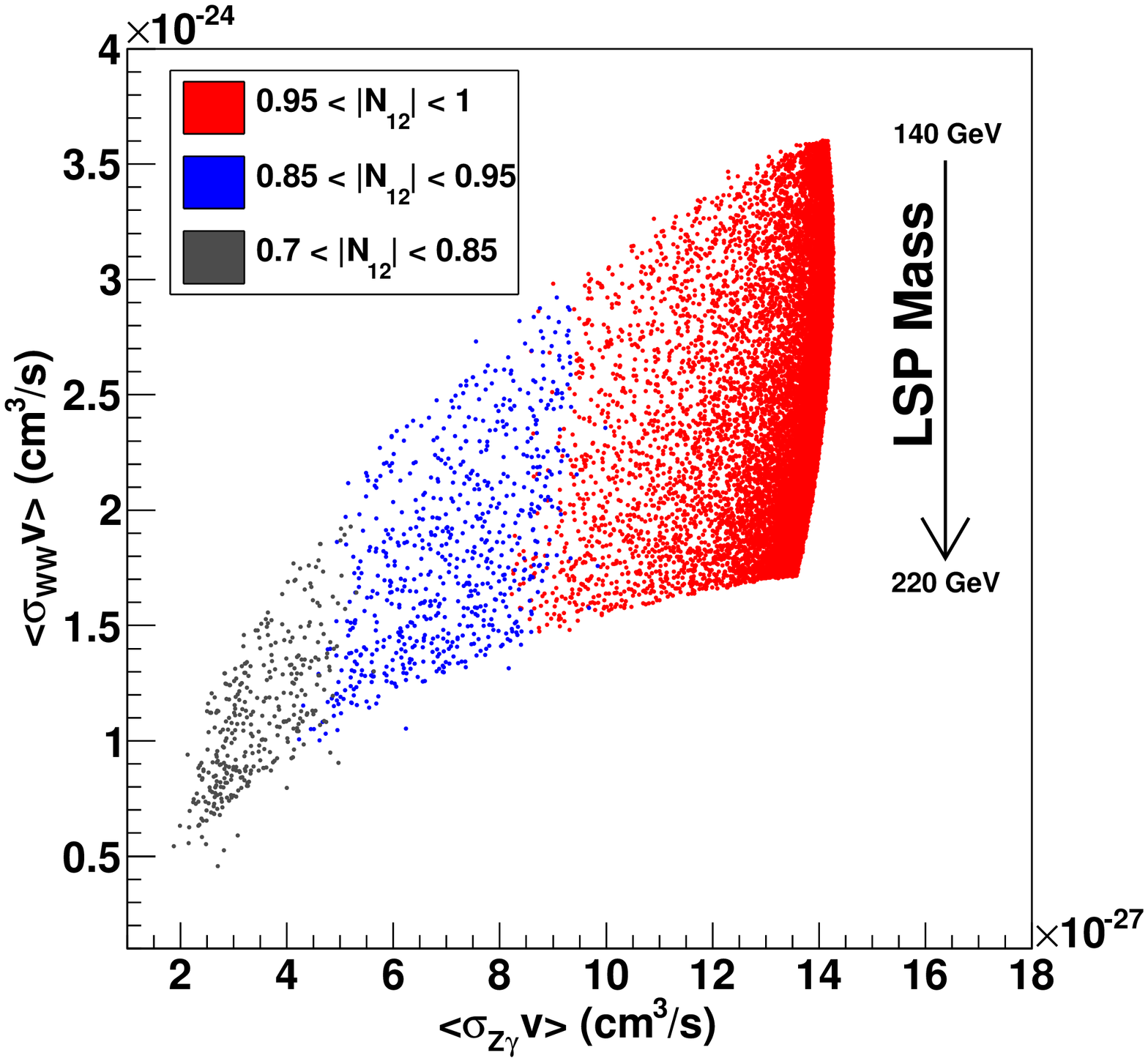}
\caption{ (Color online) Left: The models indicated by red [dark triangles] are  $G_2$ models (pure wino and decoupled scalars) and
give a good a description of the PAMELA positron data. The green [dark squares]  are
$G_2$-like models with a wino eigen-component $|N_{12}| > 0.9$ (where
the normalized LSP wino component is $N_{\tilde W} \equiv N_{12}$),
with the lightest colored superpartner being the gluino $< 700\, \rm
GeV$ but scalars can be of comparable size. Such models also describe PAMELA well.
 Grey [lighter points]  correspond to wino-like models
with wino eigen-component $|N_{12}|  > 0.7 $ with variable scalar and gluino masses. 
The upper limits are
the FERMI data with either an NFW and isothermal profile
assumed~\cite{Farnier,Murgia,Collaboration:2010ex}. Regions which
remain unconstrained by the FERMI data lie below the horizontal base of the arrows for a
given profile.
Right: Illustrating the strength of $\langle \sigma v \rangle$ from
neutralino annihilations into $WW$ relative to $\gamma
Z $  and the sensitivity of the cross sections to the wino content.
The mass of the LSP can be determined via $E_{\gamma} = M_{\rm LSP}(1 - \delta_M)$, with $\delta_M = M^2_{Z}(4 M^2_{\rm LSP})^{-1}$.
Combining the analysis of both figures shows that there
is a significant region of parameter space in wino dominated models (with an accompanying light gluino)  that are within reach
of the FERMI data and which produce cross sections in the halo that are consistent with the PAMELA data.
}
\label{figdata2}
\end{center}
\end{figure*}
In greater generality, the nature of soft breaking and
the cosmological history of the universe may very well be closely tied together \cite{Acharya:2009zt}.
On the other hand, in a non-thermal framework one
can also approach the WMAP constraint 
so long as $T_R$ does not spoil BBN constraints \cite{GelminiGondolo}.

In a thermal paradigm the relic abundance of a wino-like LSP can also be brought in
accord with the WMAP data in the presence of  residual Abelian gauge factors
that survive down to the SUSY scale and mix weakly with the MSSM neutralinos
leading to a co-annihilation enhancement \cite{Feldman:2009wv} in an otherwise depleted relic abundance from the large annihilations of the LSP.
This is to be contrasted with enhancements in the halo cross section, i.e. through a
Sommerfeld enhancement \cite{Hisano:2004ds} or through a  Breit-Wigner
enhancement \cite{Feldman:2008xs} or a boost in the flux via dark matter clumps \cite{Bergstrom:1998jj,Hooper:2003ad,Hooper:2008kv}.
Thus predictions on the relic density consistent with the production
of positrons in the halo are rather model dependant, but nevertheless can account
for the proper relic abundance of dark matter in such models.

\subsection{Connection to the Positron Data \label{positrons}}
The data released by the PAMELA collaboration  indicates a large excess in
positron flux in the halo.
For the case of models with MSSM field content,
annihilations of the LSP into $W$ bosons are dominant possible sources of positrons and indeed
the $W^+W^-$ production provides the needed cross section in the halo to account for the PAMELA anomaly
for a pure wino \cite{Kane:2009if}\cite{Feldman:2009wv} without any boost factor in the positron flux
($\langle \sigma v \rangle \sim 2.5 \times 10^{-24} ~{\rm cm^3/s}$).
The PAMELA data can  also be fit when the LSP has a non-negligible Higgsino component  \cite{Feldman:2009wv}
with small boost (clump) factors in the positron flux  $\sim 2-4$.
 Figure (\ref{figpam}) illustrates fits to the data for {various neutralino masses}
with no boost factor in the positron flux. The figure is meant
to show that models with wino-like LSPs which describe the PAMELA positron
ratio should have masses in the range near (170 - 200) GeV.
Progress has been made towards a complete fit to
both the PAMELA positron, antiproton data, and the FERMI $e^{+} +e^{-}$ flux data using {\sffamily GALPROP} \cite{Kane:2009if}
and more exhaustive analyses are currently under way.
A lighter LSP could also produce the PAMELA signal with a different set of
propagation parameters. If other effects, such as small density fluctuations are included,
the LSP mass range could cover slightly heavier masses.

\subsection{Photon Line Spectrum and Recent Probes}
The relative strength of the photon  line spectrum  arising from
dark matter annihilations in the galaxy
\cite{Bergstrom:1997fh} is highly sensitive to the
gaugino content of the LSP \cite{Ullio:2001qk}\cite{Yaguna:2009cy}.
Thus with an essentially pure wino, as in the G$_2$ models, $\langle
\sigma v \rangle^{1-\rm loop}_{\gamma Z} \sim 10^{-26}~\rm cm^3/s$,
for a wino mass corresponding to the line energy of Fig.(\ref{figdata2}).
Such models provide promising probes for dark matter candidates with
the FERMI data \cite{Collaboration:2010ex} in the central galaxy
and from dwarf galaxies. In Fig.(\ref{figdata2}) we
illustrate this effect for the recently released photon data \cite{Collaboration:2010ex}. 
The analysis shows that annihilations of a
pure wino are not inconsistent with an isothermal profile
(which may be favoured by recent simulations including baryons
\cite{Governato:2009bg}). Such a constraint is highly
dependent on the profile uncertainties. At present, the PAMELA data can be
described consistently with the FERMI {photon data 
and} Fig.(\ref{figdata2}) shows that FERMI is close to sensitivity needed
to see a signal in the line source.

Another  probe of annihilating dark matter comes from the
 FERMI analysis on dwarf galaxies  \cite{Farnier,Murgia}.
The recently reported results   
show the strongest constraints are from Ursa Minor and Draco
implying a signal should
be seen for wino masses below $\sim 300 ~\rm GeV$. This constraint
assumes a NFW dwarf density {profile}  \cite{Collaboration:2010ex}  
(see however \cite{Burkert:1995yz,Governato:2009bg,Pasetto:2010se}).  
There presently is a rather appreciable uncertainty in the predicted flux from the dwarf galaxies
due in part to the integration over the density (squared) source of dark matter \cite{Bergstrom:2005qk},\cite{Strigari:2006rd}.
For the case of
the Draco dwarf galaxy Ref. \cite{Essig:2009jx} finds an uncertainty of a factor of 10 or more.
A more detailed analysis
will help shed light on these constraints.
It would be premature to 
deduce that the constraints are ruling out models until the profile of
the dwarf galaxies are better understood and the inclusion of more stars 
enters into the analyses.

\subsection{CDMS and XENON}
A  related indication of wino-like dark matter (but not pure wino)
is that of an enhanced spin independent (SI) cross section when the
wino content is supplemented by non-negligible sources of  Higgsino
and bino content. The spin dependent cross section is also enhanced,
and their contribution is not negligible, at least for Xenon based
targets. For the SI interactions with admixtures of the above type
one finds SI cross sections in the interesting region of $\sim
{\mathcal O} (10^{-44}) ~ \rm cm^2$
\cite{Bertin:2002sq,BirkedalHansen:2002am}, (for recent related work
see \cite{DDt,nelson,DDFLN,bnelson,Hisano:2009xv,Cohen:2010gj}). For a pure wino, the tree
level cross section involving the Higgs exchanges vanish and loop
corrections \cite{Hisano:2004pv}
 are not large enough to bring
the cross section up in the region that is presently testable.
Thus observation of a signal in CDMS II, {XENON-100} (or {EDELWEISS}  and other related experiments) would immediately
exclude a pure wino LSP. Deviating from the pure wino by a few percent leads to a detectable (SI) cross section.
For example, with soft breaking parameters 
$ (m_0,m_{1/2},A_0,\tan \beta, (\Delta_1, \Delta_2, \Delta_3))
=((3000,500, 0)~ \rm GeV,4,( 0, -.56, -0.80)) $,
with ${\rm sign}(\mu)> 0$,
the LSP forms a \emph{wino-like eigenstate}:
$ (N_{\tilde{B}}, N_{\widetilde{W}},N_{\tilde{H}_1 },
N_{\tilde{H}_2})=( 0.114, -0.983, 0.127, -0.061)$,
\emph{with both a large halo annihilation cross section, $\langle \sigma v \rangle_{\w \w \to W^{+} W^{-}}$, and detectable
SI scattering cross sections, $\sigma_{\rm SI}(\w p)$.} Specifically, one obtains the {following:}
$\sigma_{\rm SI}(\w p) = 1 \times 10^{-8} ~~ {\rm pb }$, and
$\sigma_{\rm SD}(\w p)= 6 \times 10^{-6} ~~ { \rm pb }$ with
$ \langle \sigma v \rangle_{\w \w \to W^{+} W^{-}} = 2 \times
10^{-24}~{\rm cm^3/s}$. Here the LSP mass is $m_{\w} = 181~ \rm GeV $ and the gluino mass
is very light $m_{\tilde g} = 357 ~ \rm GeV$. Such a model would produce discoverable
jet signatures immediately at the LHC.
Thus, this class of model produces positrons in the halo which
describe the PAMELA data, and produces a spin independent scattering
cross section within reach of the CDMS and XENON experiments (see:
\cite{wsFK} and \cite{bnelson} for a similar emphasis). On
the other hand if XENON-100 sees no signal, and the PAMELA data
turns over at higher energies, a pure  wino remains a possible and
well-motivated interpretation.

\vspace{-.2cm}
\section{Conclusion \label{conc}}
\noindent
 In this letter we have studied  collider and dark matter implications
within the setting of soft supersymmerty
breaking based on string compactifications  and in related models with non minimal gaugino sectors.
The implications of a pure wino and a wino-like LSP in association with
the production of light gluinos at the LHC, along with  a possible
interpretation of dark matter annihilations as a cause for the rising positron ratio in PAMELA satellite data, all  provide exciting
possibilities
for the early discovery of supersymmetry. Such a discovery
will have strong implications for the underlying theory and for
the nature of soft supersymmetry  breaking, as well as for the cosmological
history of the universe.

An underlying theory which can accommodate the
positron excess, can produce testable event rates in direct
detection experiments, and lead to  testable signatures at the LHC
due to the presence of light gluinos, all can arise with an LSP that has
a substantial wino component.
In addition, the wino-like LSP can have
 a  spin independent interaction cross section 
that can be rather large when a non-negligible Higgsino component is
present.
A theory of this kind provides a compelling
candidate to explain the nature of dark matter, its relic density
from re-heating, and its annihilations in the galaxy.

Recent photon constraints from FERMI on
the above class of models are also analyzed
and we have shown that there is a large
region of parameter space where a  wino-like LSP is consistent 
with the constraints. The constraints are very sensitive to the 
gaugino content of the wavefunction of the LSP and to the assumed
halo profile. This parameter space accommodates light gluinos and
therefore jets and missing energy signals that can be tested with
early data at the LHC. 

We have particularly emphasized, via specific
models,
a light gluino, and the importance of three-body decay chains which
yield large jet multiplicities from the light gluino decays 
producing wino or  wino-like LSPs. The
resulting set of decays are strikingly simple and predictive with
gaugino production controlling the event topologies. The nearly
degenerate charginos and neutralinos arise from the three body
decays of the gluino and could be identified with a careful
analysis after collecting a sample of gluino events.
 Indeed the models discussed here are  ripe for studies at
the LHC with low luminosities and at start up center of mass
energies due to their large multi-jet event rates.

\vspace{-.4cm}
\section*{Acknowledgements}
\vspace{-.2cm}
We collectively would like to thank Bobby Acharya, Elliot Bloom, Katherine Freese, Simona Murgia, 
Aaron Pierce,  Jing Shao, Scott Watson and Kathryn Zurek for
a broad range of discussions.
This work was supported by National Science Foundation Grant PHY-0653587, and support from the Michigan
Center for Theoretical Physics (MCTP) and  Department of Energy
grant DE-FG02-95ER40899.

\end{document}